RESEARCH ARTICLE

# An Agent-Based Model Framework for Utility-Based Cryptoeconomies

Kiran Karra, Tom Mellan, Maria Silva, Juan P. Madrigal-Cianci, Axel Cubero Cortes, Zixuan Zhang *

**Abstract.** In this paper, we outline a framework for modeling utility-based blockchain-enabled economic systems using Agent Based Modeling (ABM). Our approach is to model the supply dynamics based on metrics of the cryptoeconomy. We then build autonomous agents that make decisions based on those metrics. Those decisions, in turn, impact the metrics in the next time-step, creating a closed loop that models the evolution of cryptoeconomies over time. We apply this framework as a case-study to Filecoin, a decentralized blockchain-based storage network. We perform several experiments that explore the effect of different strategies, capitalization, and external factors to agent rewards, that highlight the efficacy of our approach to modeling blockchain based cryptoeconomies.

KEY WORDS

1. Agent-Based Modeling.   2. Cryptoeconomics.   3. Digital Twin.

## 1. Introduction

Cryptoeconomics is an interdisciplinary science that combines fields such as economics, cryptography, and computer science with the goal of designing and analyzing economic incentive structures for resource allocation in decentralized systems.[1] Accordingly, cryptoeconomic systems are often used to create new forms of digital currency, utilities, and markets. Because each system has its own goals and contexts in which it is applicable, cryptoeconomic incentive structures usually need to be customized for each individual application. In addition, these systems typically show features associated with Complex Systems.[1] This means that the long-term evolution of these systems cannot be easily inferred from local changes caused by individuals, which makes the task of customizing cryptoeconomic systems to support a concrete application more difficult.

Even though cryptoeconomics is a relatively young field,[2] some work has been done to address the complexities of designing and tuning decentralised economies. An exciting new approach is to use Agent-based modeling (ABM).[3] ABM is a computational modeling technique that has been used to study a wide variety of complex systems, including social systems,[4] economic systems,[5,6] and biological systems.[7] In ABM, the system is modeled as a collection of agents, each of which has its own set of rules and behaviors. The agents interact with each other and with the environment, and the system's behavior emerges from the interactions of the individual agents.[3]

Within the cryptoeconomics space, ABM has the potential to support practitioners in three main areas:

(1) Study the cryptoeconomics and robustness of the blockchain to agent behavior. As an example, Struchkov et al.[8] used ABM to test how Decentralised Exchanges would respond to stress market conditions and front-running, while Cocco et al.[9] used ABM to analyse the mining incentives in the Bitcoin network;

(2) Explore the design space of blockchain networks. For instance, ABM has been applied to compare different token designs and their impact on prediction markets;[10]

(3) Test new features and protocols. Following the fair launch allocation from Yearn Finance, a group of researchers[11] used ABMs to examine the concentration of voting rights tokens after the launch, under different trading modalities.

This paper explores how ABM can be adapted to a particular type of decentralised system, namely utility-based decentralised networks. These networks employ their own currency to provide consumptive rights on the services or the product being offered by the network.[12] Thus, these systems mediate a marketplace of providers of a specific good and users that want to consume the good. Since the entire system depends on the good being traded, any tool attempting to model such system needs to consider how changes in utility impact the system and its agents.

Therefore, we propose a framework for applying ABMs to utility-based cryptoeconomies. Our approach is complimentary to other methods in the literature[13] and builds on the work of Zhang, et. al[14] to enable multi-scale coupling between individual microeconomic preferences and protocol specific supply dynamics. Rational users of cryptoeconomic systems will base their decisions on some aspects of the network they are involved in, which in turn affects the network. This natural feedback loop is well represented in our framework.

We apply the framework to Filecoin, a decentralised data storage network,[15] and conduct two experiments that uncover interesting aspects of the network. The first explores the agents' reward trajectories under different lending rates, while the other examines how the current cryptoeconomic mechanisms of Filecoin impact wealth distribution.

The rest of this paper is organized as follows. We begin in Section 2 by presenting the general framework for applying ABM to utility-based systems. This framework is then applied to the Filecoin network, by first developing a mathematical model of Filecoin's supply dynamics in Section 3. In Section 4, we describe the ABM that leverages this model to simulate a closed loop interaction of programmable agents within the Filecoin economy. Section 5 follows with the two experiments that showcase the utility of our ABM framework to understand the Filecoin system. We conclude in Section 6 by framing the results in the context of utility based token economies and discussing future research paths.

## 2. ABM Framework for Utility Cryptoeconomies

Agent-based models (ABMs) are a tool for modeling complex systems[3] with a high degree of granularity. They consist of two primary components which interact with each other: a) the environment, which models the system under study, and b) agents, which take actions that affect the environment.

**Formal Definition**

We define the general framework of our deterministic, discrete-time ABM as follows. Let $\mathsf{E}$ denote the set of all possible *environmental variables*. For a given time $d \in \mathbb{N}$, let $E_d \in \mathsf{E}$ denote the *environment* at time $d$, and define the set $\mathsf{E} \ni \mathscr{E}_d := E_0 \times \cdots \times E_d$. In our specific setting, $E_d, E_{d+1}, \ldots$ corresponds to the environments at day $d$, at day $d+1$, etc. In addition, let $\mathscr{A}$ be the abstract set of *agents* and let $\mathscr{H}$ denote the abstract set of actions. To each agent $a \in \mathscr{A}$, there corresponds a given set of actions $h_a \in \mathscr{H}$. Furthermore, we define the *update rules* $f_{d+1}^H : \mathscr{A} \times \mathscr{E}_d \mapsto \mathscr{H}$, $f_{d+1}^E : \mathscr{H} \times \mathscr{E} \mapsto \mathscr{E}$, and $f_{d+1}^A : \mathscr{E}_d \mapsto \mathscr{A}$, as some abstract functions that update the *actions of each agent, the environment and agents* respectively. Given a set of agents $A_0 \subset \mathscr{A}$, at time $d = 0$, where each agent $a_0 \in A_0$ is equipped with a set of actions $h_{a_0} \in \mathscr{H}$ and an initial environment $E_0$, the ABM proceeds by iterating as follows:

$$h_{a,d+1} = f_d^S(a_d, \mathscr{E}_d) \quad \forall a_d \in A_d, \tag{1}$$

$$E_{d+1} = f_d^E \left( \underset{a_d \in A_d}{\times} h_{a,d+1}, \mathscr{E}_d \right) \tag{2}$$

$$A_{d+1} = f_d^A(\mathscr{E}_d). \tag{3}$$

**Environment**

The supply dynamics, defined as factors which affect the supply of tokens in the cryptoeconomy, are modeled by the environment, $\mathsf{E}$. Factors include the total supply of tokens, the rate at which new tokens are mined, and the rate at which tokens are taken out of circulation through means such as burning. This information is often used by investors and traders who want to understand the potential value of a cryptoeconomy to then make decisions about potential investments into that economy.

Three aspects of the environment must be defined:
(1) Network Performance Metrics - what are the key set of metrics that are to be modeled and used as performance indicators when evaluating the results of the ABM simulation?
(2) Inputs - when actors take part in the cryptoeconomy, what is the subset of actions that they can take that would affect the network metrics?
(3) Outputs - what are the outputs that flow back to miners, which in turn affect miners decisions about their actions in the next time step? For example, in a typical blockchain, rewards are issued to miners and because the rate and trajectory of rewards affects agent's behavior, this is a key output. Additional outputs, such as a subset of the overall network metrics that miners can use to make rational decisions can also be included.

**Agents**

Miners in blockchains are mapped to agents, $A_i$. Miners take actions that correspond to the inputs of the environment and make those decisions based on the outputs of the environment they are interacting with. The actions they take affect the network's performance metrics, which then affects the outputs that the agents are fed. Through this feedback loop, the dynamical nature of the cryptoeconomy is modeled.

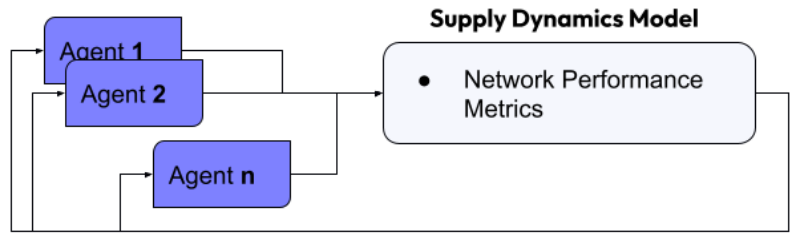

Fig. 1. Proposed framework for mapping cryptoeconomies to an ABM. Miners are represented by agents and the environment is mapped to a supply dynamics model.

Fig. 1 shows a diagram of this proposed framework for mapping cryptoeconomies to an ABM.

**Examples**

Three examples of cryptoeconomic projects where this framework can provide value include Helium,[16] Ethereum,[17] and Filecoin.[15]

Helium is a decentralized wireless network where miners provide wireless coverage in exchange for HNT tokens. An ABM utilizing the described framework can be used to understand, for example, how the rate of token distribution and population density may affect expected network coverage. This requires a mathematical model of how tokens are minted (the supply dynamics), given agent inputs (i.e. the wireless coverage they provide to the network). This can then be used to design new incentive structures to ensure a more even coverage distribution.

Ethereum is another example where the described ABM framework can be applied. A specific use case could be analyzing how user behavior (agents) could stress and effect the total circulating supply of ETH tokens after the "Shapella" upgrade, which enabled easier unlocking of staked ETH. In this setting, one could model, e.g., the propensity of a participant to unlock, against more staking inflows *due to* the ability to unlock. Here, the agents actions (whether to stake or unstake) have a direct effect on the supply dynamics and the outlined framework would enable one quantify various scenarios that may play out.

We now discuss the application of the ABM framework to Filecoin in more detail.

## 3. Modeling Filecoin Supply Dynamics

Filecoin is a distributed storage network based on the blockchain, where miners, referred to as storage providers (SPs), provide storage capacity for the network and earn units of the Filecoin cryptocurrency (FIL) by periodically producing cryptographic proofs that certify they are providing the promised storage capacity. In contrast to using Nakamoto-style proof of work to maintain consensus on the chain, Filecoin uses proof of storage: a miner's voting power — the probability that the network elects a miner to create a new block — is proportional to their current quality-adjusted storage in use in relation to the rest of the network. The cryptoeconomics of Filecoin are designed to incentivize storage providers to participate and grow the collective utility of the data storage network.

The following subsections describe various aspects of the Filecoin supply dynamics.

**Circulating Supply**

Filecoin's circulating supply $S_d$ is modeled at a daily ($d$) level of aggregation and has four parts:

$$S_{d+1} = \underbrace{M_d + V_d}_{\text{inflow}} - \underbrace{L_d - B_d}_{\text{outflow}}. \tag{4}$$

These correspond to minted block rewards $M_d$, vested tokens $V_d$, locked tokens $L_d$, and burnt tokens $B_d$.

**Power Onboarding and Renewals**

The dynamics of $M_d$, $L_d$, and $B_d$ depend on the amount of storage power onboarded and renewed in the network.

In Filecoin, storage providers (SPs) participate by onboarding power onto the network by adding a sector for a committed duration. Power is measured in units of sectors, which can be either 32 GiB or 64 GiB in size. Each sector consists of a fraction of committed capacity (CC) and verified deal data (FIL+).[18] An SPs can choose to renew CC sectors when they expire.

We model power in aggregate terms rather than at the sector level. This means that we model the network's storage power to be split into two categories: 1) CC and 2) FIL+. This approximation is valid for the granularity that we are seeking to achieve with our modeling.

Filecoin has two methods to measure the power of the network, the network's raw byte power (RBP) and the network's quality adjusted power (QAP). Network RBP is a measure of the raw storage capacity (in bytes) of the network — it does not distinguish between the kind of data that is stored on the network. For example, empty or random data stored on the network is counted the same as a widely used dataset when computing network RBP. A second measure of network power is quality adjusted power (QAP). QAP is a derived measurement that captures the amount of useful data being stored on the network. Considering the aggregated approximation discussed above, we compute the quality adjusted power, $P_d^{\text{QA}}$ of the network on day $d$ as

$$P_d^{\text{QA}} = (1 - \gamma) \cdot P_d^{\text{RB}} + 10 \cdot \gamma \cdot P_d^{\text{RB}}, \tag{5}$$

where $\gamma \in [0, 1]$ is the overall FIL+ rate of the network, and $P_d^{\text{RB}}$ is the raw byte power of the network on day $d$. Eq. 5 reveals that FIL+ power is given a 10x multiplier when computing the QA power of the network.[18]

An initial pledge collateral of FIL tokens is required in order to onboard or renew power, and the specific amounts and time-windows are discussed below. In exchange for onboarding and renewing power onto the network, and continually submitting storage proofs to the chain, SPs can receive block rewards in the form of FIL tokens.

**Rewards from minting**

Filecoin uses a hybrid minting model that has two components — simple minting and baseline minting. The total number of tokens minted by day $d$ is the sum of these two minting mechanisms:

$$M_d = M_d^{\text{S}} + M_d^{\text{B}} \tag{6}$$

Simple minting is defined by an exponential decay model

$$M_d^S = M_\infty^S \cdot (1 - e^{-\lambda d}), \tag{7}$$

which decays at a rate of $\lambda = \frac{\ln(2)}{6yrs}$, corresponding to a 6-year half-life. $M_\infty^S$ takes a value of 30% of the maximum possible minting supply of 1.1B tokens. Tokens emitted via simple minting are independent of network power. This is similar to minting schemes present in other blockchains.[19]

The second component of minting in Filecoin is baseline minting, $M_d^B$. Baseline minting depends on network power and aims to align incentives with the growth of the network's utility. The minting function still follows an exponential decay, however, it now decays based on the effective network time, $\theta_d$. The equations describing this are:

$$\begin{aligned} M_d^B &= M_\infty^B \cdot (1 - e^{-\lambda \theta_d}) \\ \theta_d &= \frac{1}{g} \ln\left(\frac{g \overline{R}_d^\Sigma}{b_0} + 1\right) \\ \overline{R}_d^\Sigma &= \sum_{d \in D} \min\{b_d, P_d^{RB}\} \end{aligned} \tag{8}$$

From these definitions, we can compute the cumulative baseline minting by day $d$ from the cumulative capped RBP of the network:

$$\begin{aligned} M_d^B &= M_\infty^B \cdot \left(1 - e^{\frac{-\lambda}{g} \ln\left(\frac{g\overline{R}_d^\Sigma}{b_0}+1\right)}\right) = \\ &= M_\infty^B \cdot \left(1 - \left(\frac{g\overline{R}_d^\Sigma}{b_0} + 1\right)^{\frac{-\lambda}{g}}\right) \end{aligned} \tag{9}$$

In this expression $M_\infty^B$ takes a value of 70% of the maximum possible supply of 1.1B tokens. $\overline{R}^\Sigma$ is the cumulative capped network RBP; it is the sum of the point-wise minimum of network's RBP and the baseline storage function for each day:

$$b_d = b_0 e^{gd} \tag{10}$$

The baseline storage function serves as a target for the network to hit to maximize the baseline minting rate. In this expression $g = \frac{\log(2)}{365}$, the baseline storage growth rate which corresponds to 2 year doubling, and $b_0 = 2.88888888 \text{EiB}$ is the initial baseline storage.

### Vesting

Vesting supply, which can contribute 0.9B tokens, is modelled daily, summing across the set of recipients $R$ as

$$V_d = \sum_{r \in R} V_{r,d}. \tag{11}$$

Different recipients have different linear vesting schedules.

### Locked tokens

Locked tokens in the network $L_d$ are made up of storage collaterals $L_d^S$ and vesting block rewards $L_d^R$ as

$$L_d = L_d^{S} + L_d^{R} \tag{12}$$

The locked balance for vesting blocked rewards is modeled as

$$L_d^{R} = 0.75 \sum_{\tau \leq d} \Delta M_{d-\tau} \cdot r_d^{R}. \tag{13}$$

where $\Delta M_d$ is daily minted rewards and $r_d^{R}$ is a vector specifying the release linear release over 180days.

The locked storage collateral is modeled as having the following dynamics

$$L_d^{S} = \sum_{\tau \leq d} \Delta L_{d-\tau}^{S} \cdot r_d^{S}. \tag{14}$$

where $\Delta L_d^{S}$ is newly locked storage collateral tokens, and $r_d^{S}$ is a vector specifying a release schedule.

Newly locked collateral tokens are given by

$$\Delta L_d^{S} = \Delta L_d^{SP} + \Delta L_d^{CP}. \tag{15}$$

where the 'storage pledge' locked tokens are

$$\Delta L_d^{SP} = \max\left(20 \cdot \Delta M_d, 0\right), \tag{16}$$

and 'consensus pledge' locked tokens are

$$\Delta L_d^{CP} = \max\left(\frac{0.3 \cdot S_d \cdot \Delta P_d^{QA}}{\max\left(P_d^{QA}, b_d\right)}, 0\right) \tag{17}$$

where $\Delta P_d^{QA}$ denotes new quality-adjusted power onboarded on day $d$.

**Burnt tokens**

Burnt tokens are modeled as consisting of termination fees $B_d^{T}$ and base fees from gas usage $B_d^{G}$ as:

$$B_d = B_d^{T} + B_d^{G} \tag{18}$$

where terminations are accounted for by aggregating agent decisions and gas fees as linearly increasing as $B_d^{G} = \beta d$ at average rate $\beta$.

## 4. ABM of Filecoin

Utilizing the framework developed in Section 2, we create an ABM of Filecoin. Storage providers (SPs) are mapped to agents, and the environment consists of the supply dynamics described in Section 3. We divide the environment into three logical modules: a) Network State, b) Forecasting, and c) the external environment. These components interact with each other in the following manner.

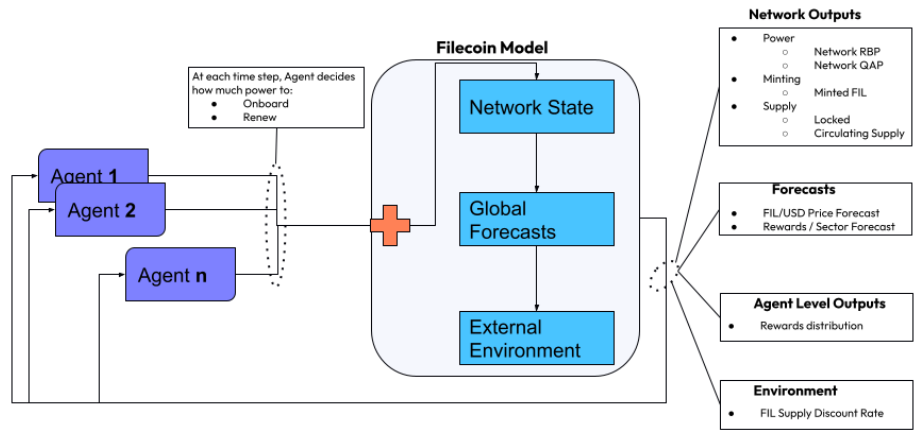

Fig. 2. Summary of the Agent Based Model of the Filecoin network, with arrows indicating direction of data flow.

Agents determine the amount of power they will onboard and renew onto the network for day $d$. All agents decisions are aggregated and passed into the network state module. Using the developed model of the supply dynamics, the network state is updated. By utilizing both historical network metrics and network forecasting information, agents can make rational decisions. Finally, an external environment simulates constraints that SPs are subject to in the real world, such as borrowing costs of pledge collateral. These components interact to create a closed-loop simulation of the Filecoin economy.

Fig. 2 summarizes the components and dataflow of the Filecoin ABM.

**Agents**

Agents directly influence the outcomes of the simulation, since their actions are aggregated and used as inputs to update the network state. We have developed three types of agents which use the network and forecasting information in different ways to make decisions regarding onboarding and renewing power:

(1) DCAAgent - This is the dollar cost averaging agent, and does not use any forecasting information or historical network information to make decisions. The agent is configured to onboard a constant amount of power per day, the percentage of that power which corresponds to verified deals, and the percentage of expiring power which should be renewed. This is a dollar-cost averaging strategy and can be useful in understanding relative performance to more complex strategies.

(2) FoFRAgent - This agent utilizes the `rewards/sector` forecast provided by the network to internally forecast the FIL-on-FIL returns (FoFR) of onboarding sectors for various sector durations, where $\text{FoFR} = \frac{\text{rewards}}{\text{pledge}}$. This metric can additionally be generalized to introduce arbitrary cost structures. Because pledge (Eq. 17) is dependent upon the Network QAP and agents must make decisions for a given timestep before the overall Network QAP is aggregated for a day, it is approximated by using the previous day's pledge. If the estimated FoFR for any of the tested sector durations exceeds a configurable threshold (which indirectly represents the risk profile of the agent), then the agent will onboard a configured amount of power. It will also renew a configured amount of power under the same condition.

(3) NPVAgent - This agent utilizes the `rewards/sector` forecast provided by the network to compute the net present value (NPV) of onboarding power for various sector durations. NPV is the present value of the expected `rewards/sector` less `costs/sector`. Present value is computed using the continuous discounting formula. The agent discount rate, configured upon instantiation, is a proxy for the risk profile of the agent, with a higher discount rate representing a higher risk aversion. The agent will onboard and renew power at a sector duration which maximizes NPV, but will take no action for day $d$ if $NPV < 0$ for all durations tested.

**Network State**

Network state consists of: a) network power, and b) the status of tokens.

The total network power is summed across all of the agents individual contributions for each day, and the type of power onboarded (CC or FIL+) is also tracked. This enables the network state to track both $P_d^{\text{RB}}$ and $P_d^{\text{QA}}$.

Token status consists of three parts: a) the amount mined, b) the amount locked due to pledge, and c) the remaining that has been released into the circulating supply through vesting. The mined tokens are distributed to agents as rewards. The fraction of total rewards mined in a day are distributed proportional to the fraction of total network QAP. Tokens are locked when agents onboard power to provide a consensus pledge. This information is aggregated to compute the overall network state of the number of tokens locked. Using Eq. 4, the circulating supply of the network is computed.

**Forecasting**

To enable agents to make rational decisions, relevant forecasts are computed and provided to agents, who can choose to utilize them. As previously mentioned, agents use `rewards/sector` to make rational decisions. This quantifies the amount of rewards an agent can expect to receive for onboarding a given sector. The metric is forecast by first utilizing the historical network RBP and QAP to train a time-series forecasting model, $\mathcal{M}$. $\mathcal{M}$ is used to forecast RBP and QAP trajectories until the end of the simulation, denoted by $\hat{P}_d^{\text{RB}}$ and $\hat{P}_d^{\text{QA}}$, respectively. $\hat{P}_d^{\text{RB}}$ is then used to compute the expected minting rate, $\hat{m}_d$, using Eqs. 6, 7, and 9. Finally, because rewards are distributed proportional to the network's QA power, $\widehat{\frac{rewards}{sector}}[d] = \hat{m}_d / \hat{P}_d^{\text{QA}}$.

There are currently two models implemented for RB and QA forecasting, linear extrapolation and a variant of Markov-Chain Monte Carlo.[20,21] Agents may also elect to perform custom forecasting using network metrics, and this represents a competitive advantage that a certain agent may have.

**External Environment**

Agents must borrow tokens in order to satisfy the consensus pledge (Eq. 17) needed to onboard power. The borrowing rate is modeled as an external environment process that specifies the discount rate, $R_d$, at which agents can borrow tokens. Agents use this information to make rational decisions regarding onboarding and renewing power. The purpose of this is to model realities that SPs have to face, when determining their strategy for being involved in the Filecoin network. Additional real-world complexities can also be modeled here.

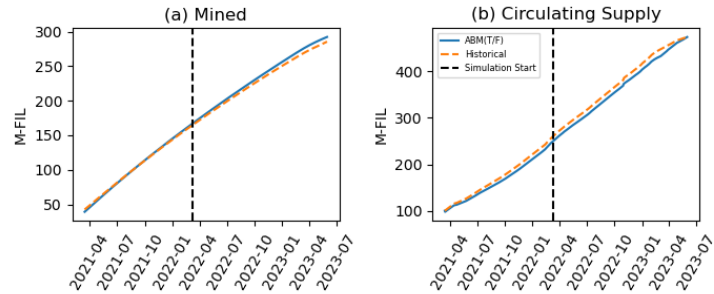

Fig. 3. Validating our supply dynamics and ABM through backtesting. (a) The mined FIL of the model to the historical data, and (b) The circulating supply computed by the model against historical data.

**Model Validation**

We begin by validating the model of Filecoin's supply dynamics that was developed and described in Section 3, using backtesting. Our approach is to instantiate one DCAAgent which onboards and renews the historical power that was onboarded onto the network for that day. This is in contrast to the typical use-case of an agent, which is making daily decisions about whether and how much power to onboard and renew. Then, the relevant statistics for circulating supply are calculated from the start of simulation to the present date. This is then compared against actual statistics from the Filecoin network retrieved from Spacescope.[22] Fig.3 shows the results of this experiment: a) shows the minted tokens, and b) shows the circulating supply. For each of these network statistics, the implemented model tracks the historical data with good accuracy. Slight differences are observed, and these can be attributed to not modeling certain intricacies of the Filecoin network, such as variable sector durations.

## 5. Experiments and Results

In this section, we describe some experiments that showcase the utility of ABM in modeling blockchain networks, using the Filecoin network as a case study.

**Sensitivity of Rewards to External Discount Rates**

In this experiment, we explore how the cryptoeconomics of Filecoin and external factors such as borrowing rates affect agent rewards. We instantiate two subpopulations of NPVAgents, one subpopulation is configured to only onboard verified deals (which corresponds to FIL+ power), while the second is configured to only onboard storage capacity (CC power). Both subpopulations of agents are configured to have identical risk profiles by instantiating the agents with the same discount rates. Fig. 4 shows agent rewards trajectory, with different colors indicating different external discount rates.

Fig. 4(a) shows that irrespective of external discount rates, FIL+ agents are more profitable than CC agents. This is a direct consequence of the cryptoeconomic mechanism in place in

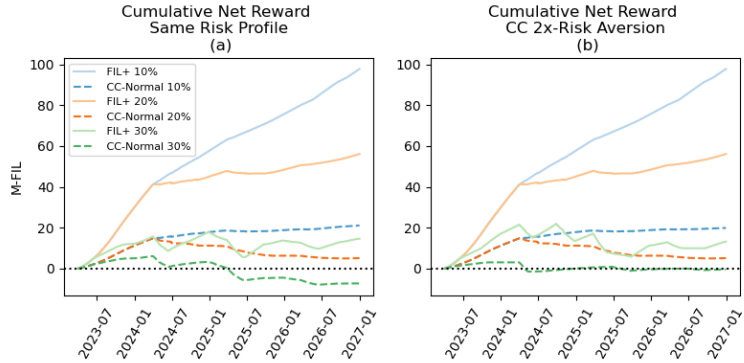

Fig. 4. Experiment exploring the sensitivity of returns to external discount rates with two subpopulations of agents, FIL+ and CC. In (a) both exhibit the same risk profile, (b) the CC agent has 2x the risk aversion of the FIL+ agent.

Filecoin to incentivize FIL+ data, through the 10x quality adjusted (QA) multiplier. Secondly, we see the effect of external borrowing rates on agent profitability. As expected, higher rewards are correlated with lower borrowing rates. However, the rewards trajectory does not change linearly with the borrowing rate and starts to oscillate as borrowing rates increase. This is an example of an interesting dynamic that emerges as a result of the agent based simulation.

We extend this experiment by altering one aspect of the previous experimental setup - that is, we increase the risk aversion of the CC agent to be two-times the risk aversion of the FIL+ agent. We then examine the agent rewards trajectory, shown in Fig. 4(b). Because the FIL+ agents have the same risk as before, their rewards trajectories are identical. However, we notice that when the external discount rate is 30%, the risk-averse CC agent manages a more positive rewards trajectory than the non risk-averse CC agent. The effect of this disappears as the external borrowing rates decrease, however.

**Wealth Concentration**

In this experiment, we explore how the distribution of starting capital in the cryptoeconomic network affects the ability to get rewards from the network. Our experimental setup consists of five DCAAgents which are configured to represent different levels of capitalization. This is represented with a vector $[a_1, a_2, a_3, a_4, a_5]$, where the relative capitalization of $a_i$ is defined as $c_i = a_i / \sum_{i=1}^{5} a_i$.

In Filecoin, onboarding power requires, in addition to pledge collateral, sealing of sectors via cryptographic proofs that require large computational resources. It is reasonable to assume that agents with larger capitalization will have more hardware resources to perform this than agents with smaller capitalization, thereby having a larger sealing throughput. To model this, we scale how much power an agent is able to onboard and renew, per day, by its relative capitalization. The mapping from capitalization to sealing throughput captures the idea of wealth concentration. To compare and interpret the results, the overall power onboarded is kept constant across the three experiments.

We test three distributions of initial capital:
(1) All agents have equal starting capital (20%) - this is considered the baseline and corre-

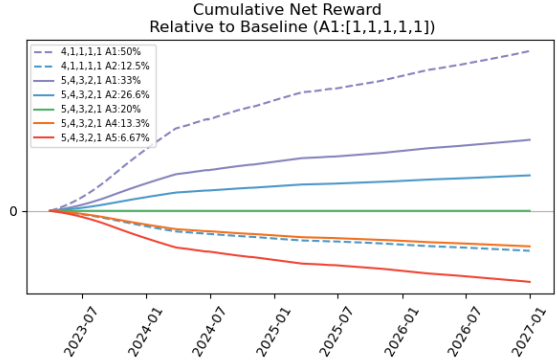

Fig. 5. The trajectory of rewards for agents with various starting capitalizations, relative to the baseline distribution, where all agents are equally capitalized (20%).

sponds to the the vector $[1, 1, 1, 1, 1]$.
(2) One agent has 50% of the starting capital, and the remainder have $50/4 = 12.5\%$ starting capital each and corresponds to the vector $[4, 1, 1, 1, 1]$.
(3) The agent capitalization follows the distribution: $[33\%, 27\%, 20\%, 13\%, 7\%]$ and corresponds to the configuration vector $[5, 4, 3, 2, 1]$.

Fig. 5 shows the reward trajectories of each agent, relative to the baseline case where each agent has 20% of the starting capital. We observe that relative to the max-capitalized agent, the rewards trajectories of other agents are on a decreasing trend. This is a consequence of the fact that both onboarding and renewals are a function of the agent capitalization.

## 6. Conclusion

In this paper, we have outlined a framework for applying ABM to modeling utility based blockchain economies, and validated our framework with Filecoin as a case study. Our experiments shed light on some interesting aspects of Filecoin, including agent reward trajectories when taking into account external lending rates, and how the cryptoeconomic structure of Filecoin distributes wealth.

The sensitivity experiments indicate that creating new, competitive lending markets with smart contracts leveraging programmable platforms such as FVM[23] can enable network growth and increase miner returns. The wealth concentration experiments indicate that starting capitalization has a significant effect on total rewards in the future. By explicitly modeling this effect with the supply dynamics, one can then design new incentive structures to either accentuate, maintain, or perhaps reverse the trend based on the goals of the project. Insights such as these, enabled by the ABM framework, can help designers and creators of cryptoeconomies to more efficiently achieve their goals. This indicates that ABM can be a valuable tool for researchers to better understand and design blockchain economies.

In the future, we plan to explore additional aspects of blockchain economies that are well mapped to ABMs, such as the effect of information quality, availability, and lag on agent reward trajectories, and related network science questions. Another potential research direction is to include uncertainty by considering a probabilistic ABM, while balancing the computational constraints using methods such as Multi-level and Multi-Index Monte Carlo methods.[24–27]

## Author Contributions

KK developed the ABM codebase and helped devise experiments that were conducted with the framework. TM developed the mathematical models of the Filecoin economy and steered the project. They both contributed equally to manuscript preparation. TM and MS implemented the initial mathematical models, which were then ported to the ABM framework. JC provided formalism and consulting on ABM related topics. AC and ZZ helped putting the project in larger context and helped with manuscript preparation.

## Notes and References

[1] Voshmgir, S., Zargham, M., *et al.* "Foundations of cryptoeconomic systems." *Research Institute for Cryptoeconomics, Vienna, Working Paper Series/Institute for Cryptoeconomics/Interdisciplinary Research* **1**.

[2] Davidson, S., De Filippi, P., Potts, J. "Economics of Blockchain." (2016) doi:10.2139/ssrn.2744751 URL https://papers.ssrn.com/abstract=2744751.

[3] Macal, C. M., North, M. J. "Agent-based modeling and simulation." In *Proceedings of the 2009 winter simulation conference (WSC)* IEEE 86–98 (2009) .

[4] Terano, T. "A Perspective on Agent-Based Modeling in Social System Analysis." In G. S. Metcalf, K. Kijima, H. Deguchi (Eds.), *Handbook of Systems Sciences* Singapore: Springer 1–13 (2020)doi:10.1007/978-981-13-0370-8_5-1 URL https://doi.org/10.1007/978-981-13-0370-8_5-1.

[5] Chen, S.-H., Chang, C.-L., Du, Y.-R. "Agent-based economic models and econometrics." *The Knowledge Engineering Review* **27.2** 187–219 (2012) doi:10.1017/S0269888912000136 publisher: Cambridge University Press URL https://www.cambridge.org/core/journals/knowledge-engineering-review/article/abs/agentbased-economic-models-and-econometrics/DF3E4987809567A9B277F83ED6E22E00.

[6] Fagiolo, G., Guerini, M., Lamperti, F., Moneta, A., Roventini, A. "Validation of Agent-Based Models in Economics and Finance." In C. Beisbart, N. J. Saam (Eds.), *Computer Simulation Validation: Fundamental Concepts, Methodological Frameworks, and Philosophical Perspectives* Cham: Springer International Publishing Simulation Foundations, Methods and Applications 763–787 (2019)doi:10.1007/978-3-319-70766-2_31 URL https://doi.org/10.1007/978-3-319-70766-2_31.

[7] Eubank, S., *et al.* "Modelling disease outbreaks in realistic urban social networks." *Nature* **429.6988** 180–184 (2004).

[8] Struchkov, I., Lukashin, A., Kuznetsov, B., Mikhalev, I., Mandrusova, Z. "Agent-Based Modeling of Blockchain Decentralized Financial Protocols." In *2021 29th Conference of Open Innovations Association (FRUCT)* 337–343 (2021) doi:10.23919/FRUCT52173.2021.9435601 iSSN: 2305-7254.

[9] Cocco, L., Tonelli, R., Marchesi, M. "An Agent Based Model to Analyze the Bitcoin Mining Activity and a Comparison with the Gold Mining Industry." *Future Internet* **11.1** 8 (2019) doi:10.3390/fi11010008 number: 1 Publisher: Multidisciplinary Digital Publishing Institute URL https://www.mdpi.com/1999-5903/11/1/8.

[10] Hülsemann, P., Tumasjan, A. "Walk this Way! Incentive Structures of Different Token Designs for Blockchain-Based Applications." *ICIS 2019 Proceedings* URL https://aisel.aisnet.org/icis2019/blockchain_fintech/blockchain_fintech/7.

[11] Fernandez, J. D., Barbereau, T., Papageorgiou, O. "Agent-based Model of Initial Token Allocations: Evaluating Wealth Concentration in Fair Launches." (2022) 2208.10271.

[12] Benedetti, H., Abarzúa, L., Caceres Fuentes, C. "Utility Tokens." (2021) doi:10.2139/ssrn.4088568 URL https://papers.ssrn.com/abstract=4088568.

[13] Akcin, O., Streit, R. P., Oommen, B., Vishwanath, S., Chinchali, S. https://eprint.iacr.org/2022/1492 "A Control Theoretic Approach to Infrastructure-Centric Blockchain Tokenomics." (2022) Cryptology ePrint Archive, Paper 2022/1492 URL https://eprint.iacr.org/2022/1492.

[14] Zhang, Z., Zargham, M., Preciado, V. M. "On modeling blockchain-enabled economic networks as stochastic dynamical systems." *Applied Network Science* **5.1** 1–24 (2020).

[15] Benet, J., Greco, N. "Filecoin: A decentralized storage network." *Protoc. Labs* 1–36.

[16] Haleem, A., Allen, A., Thompson, A., Nijdam, M., Garg, R. "A decentralized wireless network." *Helium Netw* 3–7.

[17] Buterin, V., *et al.* "A next-generation smart contract and decentralized application platform." *white paper* **3.37** 2–1 (2014).

[18] Labs, P. apr 28, 2023 "Filecoin Spec." (2018) URL https://spec.filecoin.io.

[19] Nakamoto, S. "Bitcoin: A peer-to-peer electronic cash system." *Decentralized business review* 21260.

[20] Neal, R. M., *et al.* "MCMC using Hamiltonian dynamics." *Handbook of markov chain monte carlo* **2.11** 2 (2011).

[21] Hoffman, M. D., Gelman, A., *et al.* "The No-U-Turn sampler: adaptively setting path lengths in Hamiltonian Monte Carlo." *J. Mach. Learn. Res.* **15.1** 1593–1623 (2014).

[22] Labs, S. apr 28, 2023 "Spacescope." (2018) URL https://spacescope.io.

[23] Accessed: 2022-06-09 "Introducing the Filecoin Virtual Machine." https://filecoin.io/blog/posts/introducing-the-filecoin-virtual-machine/.

[24] Giles, M. B. "Multilevel monte carlo methods." *Acta numerica* **24** 259–328 (2015).

[25] Madrigal-Cianci, J. P., Kristensen, J. "Time-efficient Decentralized Exchange of Everlasting Options with Exotic Payoff Functions." In *2022 IEEE International Conference on Blockchain (Blockchain)* 427–434 (2022) doi:10.1109/Blockchain55522.2022.00066.

[26] Madrigal-Cianci, J. P., Nobile, F., Tempone, R. "Analysis of a class of multilevel Markov chain Monte Carlo algorithms based on independent Metropolis–Hastings." *SIAM/ASA Journal on Uncertainty Quantification* **11.1** 91–138 (2023).

[27] Qian, E., Peherstorfer, B., O'Malley, D., Vesselinov, V. V., Willcox, K. "Multifidelity Monte Carlo estimation of variance and sensitivity indices." *SIAM/ASA Journal on Uncertainty Quantification* **6.2** 683–706 (2018).